\documentclass[sigconf]{acmart}
\AtBeginDocument{%
 \providecommand\BibTeX{{%
 \normalfont B\kern-0.5em{\scshape i\kern-0.25em b}\kern-0.8em\TeX}}}

\usepackage{enumitem}
\usepackage{multirow}
\usepackage{makecell}

\newlist{questions}{enumerate}{2}
\setlist[questions,1]{label=RQ\arabic*.,ref=RQ\arabic*}
\setlist[questions,2]{label=(\alph*),ref=\thequestionsi(\alph*)}

\copyrightyear{2024} 
\acmYear{2024} 
\setcopyright{rightsretained} 
\acmConference[CHI '24]{Proceedings of the CHI Conference on Human Factors in Computing Systems}{May 11--16, 2024}{Honolulu, HI, USA}
\acmBooktitle{Proceedings of the CHI Conference on Human Factors in Computing Systems (CHI '24), May 11--16, 2024, Honolulu, HI, USA}
\acmDOI{10.1145/3613904.3642061}
\acmISBN{979-8-4007-0330-0/24/05}

\begin{document}

\title{Firefighters' Perceptions on Collaboration and Interaction with Autonomous Drones: Results of a Field Trial}


\author{Moyi Li}
\email{moyi.li@uzh.ch}
\orcid{0009-0008-9638-7890}
\affiliation{%
 \institution{University of Zurich}
 \streetaddress{}
 \city{Zurich}
 \state{}
 \country{Switzerland}
 \postcode{}
}

\author{Dzmitry Katsiuba}
\email{dzmitry.katsiuba@uzh.ch}
\orcid{0000-0002-4341-5738}
\affiliation{%
 \institution{University of Zurich}
 \streetaddress{}
 \city{Zurich}
 \state{}
 \country{Switzerland}
 \postcode{}
}

\author{Mateusz Dolata}
\email{mateusz.dolata@uzh.ch}
\orcid{0000-0002-2732-4465}
\affiliation{%
 \institution{University of Zurich}
 \streetaddress{}
 \city{Zurich}
 \state{}
 \country{Switzerland}
 \postcode{}
}

\author{Gerhard Schwabe}
\email{gerhard.schwabe@uzh.ch}
\orcid{0000-0002-0453-9762}
\affiliation{%
 \institution{University of Zurich}
 \streetaddress{}
 \city{Zurich}
 \state{}
 \country{Switzerland}
 \postcode{}
}


\begin{abstract}
 Applications of drones in emergency response, like firefighting, have been promoted in the past decade. As the autonomy of drones continues to improve, the ways in which they are integrated into firefighting teams and their impact on crews are changing. This demands more understanding of how firefighters perceive and interact with autonomous drones. This paper presents a drone-based system for emergency operations with which firefighters can interact through sound, lights, and a graphical user interface. We use interviews with stakeholders collected in two field trials to explore their perceptions of the interaction and collaboration with drones. Our result shows that firefighters perceived visual interaction as adequate. However, for audio instructions and interfaces, information overload emerges as an essential problem. The potential impact of drones on current work configurations may involve shifting the position of humans closer to supervisory decision-makers and changing the training structure and content. 
\end{abstract}



\begin{teaserfigure}
 \frame{\includegraphics[width=\textwidth]{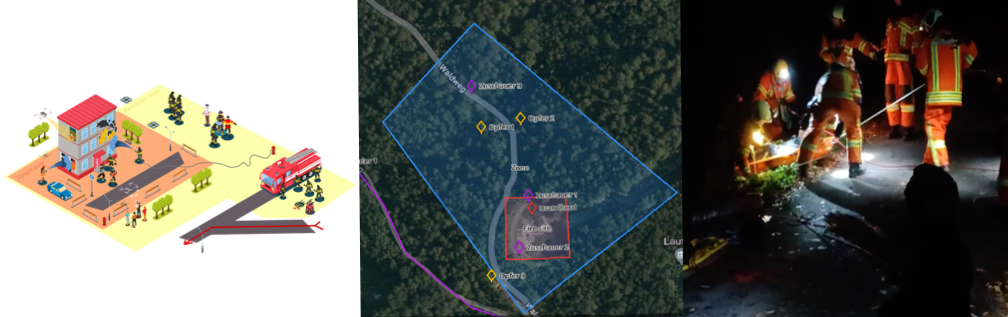}}
\end{teaserfigure}


\maketitle

\section{Introduction}
Introducing new technologies catalyzes a transformation in traditional system operation models. In the area of emergency response, drones are especially anticipated as a tool that, through their mobility, could provide new information, shorten response time, or perform dangerous tasks. Drones have also proven practical and reliable for such tasks \cite{roldan-gomez_survey_2021,ausonio_drone_2021}. In the foreseeable future, such aerial vehicles might become integral to many fire departments worldwide. However, their integration in firefighting processes and their acceptance by the public remain a challenge \cite{dolata_moving_2023}. Specifically, it is unclear how existing processes need to be adapted to incorporate drone system.

Firefighters often confront a hazardous working environment in searching, rescuing, and fire-extinguishing. Despite fire brigades' current efforts to enhance safety measures to minimize casualties \cite{kang_analysis_2016,smith_firefighter_2011}, there remain numerous unpredictable factors during their operations that can lead to accidents \cite{khan_review_2022}. Hence, as technology permits, the idea of using drones to complete specific hazardous tasks has emerged, offering additional options for conducting firefighting operations. The applications of drones are gradually extending beyond tasks that endanger the health of firefighters: drones are used to collect additional data that were not available before, explore the surroundings, or assure the safety of the area. This prompts scholars to contemplate the potential for variability in the roles of drones within firefighting teams and the potential implications for current organizational structures. Exploratory research \cite{khan_exploratory_2019} shows that firefighters and emergency responders need precise information and reliable communication and have concerns about drones' ability to deal with emergencies and make decisions. Domain experts also identify the design needs for the interface of drone systems in emergencies, such as distinct visual cues and explanations for drone decisions \cite{agrawal_next_2020}. Another study proposes the direct communication between firefighters and drones, and presents a set of gestures \cite{alon_drones_2021}. Results of \cite{dolata_morphological_2022} point out the need to explore such collaboration from a social-technical perspective. Later work identifies the challenges of collaboration, intermediary devices, and information extraction from videos and re-emphasizes the importance of trust \cite{hoang_challenges_2023}. However, previous studies relied primarily on surveys and interviews and did not expose firefighters or citizens sufficiently to actual operations involving drones. Since it is likely that participation in an operation might raise new issues or turn concerns obsolete, we decided to explore firefighters' and the public's concerns in a trial.

The application of drones in firefighting has been proven feasible; for instance, drones equipped with infrared sensors are already used for fire source detection \cite{christensen_use_2015, zhang_evaluating_2021, abdelrahman_see_2017}. This suggests that drones will become more prevalent in firefighting operations as technology matures. \emph{Human-drone interaction} (HDI) and its application in emergency response often involve applying knowledge from multiple fields such as philosophy, design, and engineering \cite{baytas_ihdi_2020}. The complexity brought by this makes the design of such systems more intricate, and it is unclear if prior knowledge is still applicable. Additionally, it is important to validate existing organizational frameworks in practice, i.e., we need to understand how drones fit and extend processes and hierarchies involved in firefighting operations. The importance of adequate organizational frameworks in emergency situations has been broadly discussed in earlier research. The seminal work of Weick \cite{weick_collapse_1993} analyzed the Mann Gulch fire disaster that resulted in the death of 13 firefighters. This work identified distrust in the commander's decisions, late sensemaking, and poor communication as the primary reasons for the tragic deaths of firefighters in the Mann Gulch incident. The critical question posed in Weick's research is: Is there an appropriate firefighting organizational framework to manage the hazards brought about by change \cite{weick_collapse_1993}? He believed constructing a virtual model capable of providing an overall situation and assigning roles to participants could be a potential solution. In this framework, interactions among individuals with mutual respect and the creation of a trusting environment are particularly crucial for handling unforeseen situations. We believe that drones can provide assistance in constructing the overall scenario for their ability to gather comprehensive information within a short time. Drone systems could be a suitable solution for connecting and organizing firefighters. However, the design of such systems depends on how firefighters perceive the system's functionalities.

Therefore, in this paper, we address the following two research questions:

\begin{questions}
 \item What would be the potential design of an autonomous drone system to support firefighting emergency response?\\
 \item What challenges does the use of autonomous drones pose for human-drone interaction during firefighting operations?
\end{questions}

To address our research questions, we reviewed the literature about the support provided by drones and ways of HDI in firefighting. The results of previous studies show that effective interaction between drones and humans requires a multisensory approach \cite{cacace_multimodal_2016}. Based on this, we designed an interactive drone system consisting of two drones incorporating sound, visuals, and interfaces. We tested this system in two real-life firefighting exercises. Subsequently, we conducted post hoc interviews with firefighters and bystanders and performed a qualitative analysis of the results to assess the system’s impact on existing firefighting practices.

Our results demonstrate that the autonomous drone system provides the advantage of exploring the environment more efficiently than the original organizational framework without drone support. However, as our results indicate, use of drones might create information overload and pose privacy concerns when conveying information through sounds or videos. Firefighters who participated in the trials believed that instantaneous interaction would be less efficient because it occupies their memory space during emergent situations. Both commanders and firefighters believed that information could reach all members of the firefighting team instead of being forwarded only to the commanders. While there is no doubt about the drones' capability to perform instrumental tasks, we still consider it necessary for humans to supervise the decisions made by the drones. Introducing drones may also stimulate the need to train higher-level team personnel in robotics knowledge.

\section{Background}
\subsection{Autonomous Drones}
The drone evolution is heading toward a higher degree of autonomy. A literature review in 2015 \cite{floreano_science_2015} categorized drone autonomy into three levels: motor, reactive, and cognitive autonomy. In a 2020 study by Nonami \cite{nonami_present_2020}, the researcher further classified the current and future autonomy of drones into six levels, with the highest level signifying drones flying independently to a destination relying solely on vision. The automation of drones is now evident in functions such as automatic navigation, landing, charging, target object localization, and tracking \cite{sani_automatic_2017, woo_auto_2017, henry_automatic_2020}. Changes in autonomy may impact how people perceive interactions with drones. However, research on autonomous drones has less frequently considered how to measure and implement such changes in the design of drone systems.

Exploring autonomous drone applications across various domains represents a trending and expansive research avenue. Autonomous drones find practical applications in fields like the military, mining, photography, logistics, agriculture, and more \cite{patil_survey_2020, jones_applications_2020, mademlis_challenges_2018, budiharto_review_2019}. The results of these studies indicate that autonomous drones can play a significant role in collecting and transmitting information and assisting in rescue missions. This suggests that autonomous drones also hold great potential for applications in the field of firefighting. However, research on drones in firefighting remains relatively limited. Researchers have identified tasks where autonomous drones could potentially aid in fire source identification and detection \cite{yadav_deep_2020}. The design requirements for autonomous drone applications in firefighting and firefighters' concerns regarding autonomous drones are still under investigation. For this reason, our study includes the exploration of how firefighters perceive autonomous drones used in their work. We provide an interactive interface with events identified and decisions made by drones.

\subsection{Human-drone interaction}
To establish a system where drones assist firefighters in extinguishing fires, we have also reviewed the existing literature for proposed modes of human-drone interaction. Table \ref{tab:interactions} presents a list of interaction modes and the literature sources introducing them. Tezza and Andujar's survey \cite{tezza_state---art_2019} provides a detailed overview of various approaches to interfacing with drones. Their research encompasses controlling the system through a remote controller, gestures, voice commands, brain-computer interaction, touch input, and multimodal approaches \cite{tezza_state---art_2019}, along with their corresponding advantages and disadvantages. Gesture, voice, BCI, and touch interactions are particularly relevant to our study, as they enable direct interaction between firefighters and drones. Additionally, light appears to be another interaction method between humans and drones in spacious, open environments \cite{herdel_above_2022, abtahi_drone_2017}. 
\begin{table*}
 \caption{List of Human-drone interaction approaches in previous research}
 \label{tab:interactions}
 \begin{tabular}{ccl}
 \hline
 \textbf{Interactions} & \textbf{Literature} & \textbf{Description}\\
 \midrule
 Light & \cite{e_drone_2017} \cite{ginosar_at_2023} & Drones use light on them to send messages to humans at an outstanding position in the air. \\ \\
 Audio & \cite{mohsan_role_2022} \cite{choutri_multi-lingual_2022} & Humans and drones communicate with each other through voices and sounds.\\ \\
 Gesture & \cite{cauchard_drone_2015} \cite{medeiros_human-drone_2020} & Humans give instructions to drones by changing the shape and position of hands and arms.\\ \\
 BCI & \cite{saeedi_adaptive_2016}\cite{sanna_bari_2022}\cite{wolf_brain-computer_2022} & Humans directly communicate with drones through devices that can capture their brain movements. \\ \\
 Tactile & \cite{abtahi_drone_2017} & Humans give instructions to drones through physical contact.\\
 \hline
\end{tabular}
\end{table*}
Gesture interaction primarily involves conveying instructions to drones using different hand shapes and positions or other parts of the human body \cite{cauchard_drone_2015, medeiros_human-drone_2020}. Research on sound-based interaction has matured and gained recognition in practical applications. Common sound-based interaction is drones transmitting information to humans in a unidirectional way \cite{mohsan_role_2022}, while literature \cite{choutri_multi-lingual_2022} also introduced the bidirectional voice interaction through interface or controller. \emph{Brain-computer interaction} (BCI) is an approach that enables direct communication between the human brain and digital devices. This type of human-drone interaction in firefighting typically involves capturing human thoughts through devices and transmitting them to drones \cite{saeedi_adaptive_2016, wolf_brain-computer_2022}. It is considered a potential method for swiftly conveying human intentions to drones. Tactile interaction refers to interaction between humans and drones through direct physical contact. Light-based interaction \cite{cauchard_drone_2015, abtahi_drone_2017} typically involves drones transmitting simple messages by controlling the integrated spotlights (color, target of exposure, dynamics, etc.). The design needs of interfaces for drone systems in terms of information filtering, retrieval, and explanations are also widely discussed \cite{jones_rescuecastr_2022, hoang_challenges_2023, agrawal_next_2020}.

The above studies were primarily conducted in general human-drone interaction scenarios rather than in the context of emergency response. The potential impact of transitioning from traditional communication methods is also an open topic that requires examination. For these reasons, we designed our drone system (Section \ref{Drone_system}) based on previous interaction approaches to find out which tasks can be supported and how firefighters may react. 

\subsection{Drones in emergency firefighting response}
Researchers have been studying the application of digital technologies in firefighting operations for decades. Research in the field of firefighting technologies encompasses digital instruction files \cite{garcia-hernandez_graphical_2019}, radio communication \cite{denef_rigid_2011}, aerial and ground robots \cite{zhao_design_2022,jindal_design_2021}, intelligent protective clothing \cite{etal_study_2021, choi_designing_2021}, and organizational systems for evaluating tasks and optimizing processes \cite{jiang_ubiquitous_2004, monares_mobile_2011}. Among these, drones are considered to be particularly helpful for firefighting missions in hazardous or large areas. Engineering researchers are currently exploring drones in firefighting as a trending topic. 

Drones have demonstrated their capacity to address the limitations of human crews in extreme and hazardous environments during firefighting operations \cite{kahiluoto_supplying_2020}, as well as the crucial issues in search and rescue, such as communication and time-saving \cite{desjardins_collaboration_2014,jones_remote_2020, mayer_drones_2019}. Researchers have extensively investigated the use of drones for searching and rescuing victims \cite{garcia-hernandez_graphical_2019, pelosi_improved_2017, alharthi_activity_2021}. In addition to surveying vast or perilous areas, drones also find applications in firefighting for extinguishing fires \cite{barua_design_2020, aydin_use_2019} and improving communication signals \cite{monares_mobile_2011}. 

The utilization of multi-drone systems in firefighting has gained prominence in recent years, particularly in the context of forest fires. A survey \cite{roldan-gomez_survey_2021} examining the current deployment of drone swarms in forest fires revealed that firefighters tended to allocate drones for information collection, such as monitoring fires and assessing their potential spread risks. Sherstjuk et al. \cite{sherstjuk_forest_2019} introduced a drone swarm as a novel approach for responding to forest fires, while Ausonio et al. \cite{ausonio_drone_2021} proposed a framework for organizing drone swarms more effectively in wild firefighting, adapting to changing water needs in various scenarios. 

The potential applications of drones in firefighting appear limitless, yet there are differing opinions on their use in this context. Two studies indicated that German firefighters still harbored concerns about the limited capabilities and applications of drones in firefighting \cite{schlauderer_new_2016, weidinger_is_2018}. Past research has extensively explored the specific applications of drones in firefighting. However, there is limited literature on multi-drone systems regarding task allocation and their impact on firefighting teams. Thus, in our research, we attempt to study how an autonomous multi-drone system can be deployed and contribute to enhancing communication in firefighting through a two-role drone system providing support for scouting, evacuation, and rescue. 

\subsection{HDI in emergency response}
Human-drone interaction in emergency response has concentrated on interaction techniques, targets, communication channels, and impact on stakeholders. 911 callers and Canadian firefighters articulated their expectations and concerns regarding integral drone systems employed for communication, filming, and indoor searching, assessing whether the public would find such systems credible and trustworthy \cite{khan_exploratory_2019}. Later study investigated the design requirements for rescue system interfaces among domain experts \cite{agrawal_next_2020} and introduced and evaluated a set of gestures for interactions between firefighters in action and drones \cite{alon_drones_2021}. Recent work that delved into drone swarm system interfaces revealed that firefighters desire components to control the level of autonomy and simpler means of interaction to replace gestures and keypad input \cite{hoang_challenges_2023}. An examination of drone swarms in forest fires \cite{bjurling_drone_2020} demonstrated dynamic human-drone interaction occurring at corresponding levels in both teams, inspiring future research to explore flexible and efficient interactions between drone swarms and firefighter crews. Prior research also studied how new tools like drones could influence teamwork for search and rescue \cite{mencarini_becoming_2023}. The work of Jones et al. proposed that drones could be an approach to enhance coordination during rescue \cite{jones_remote_2020}. Social acceptance is frequently discussed in HDI \cite{oltvoort_i_2019, lidynia_droning_2017, herdel_public_2021, pytlikzillig_drone_2018, chang_spiders_2017}, especially when talking about autonomous machines \cite{tolmeijer_capable_2022}. Among these discussions, drones for public services, such as policing and firefighting, receive more trust \cite{lidynia_droning_2017, herdel_public_2021}. Still, the latest research presents the public's concerns about fairness during interactions when applying drones in police operations \cite{dolata_moving_2023}.

Despite the insightful results from the above research, we still have doubts about their reliability while applying drones in practical firefighting operations. Aligned with previous research, we believe it is important to investigate from a sociotechnical perspective when designing drone systems to identify potential needs \cite{dolata_morphological_2022}. Specifically, this refers to investigating more deeply into firefighters' reactions toward drones under firefighting. Unpredictable reactions in firefighting can cause serious, even life-threatening, consequences when the change is too dramatic. Most of the above studies were conducted in simulated settings, where firefighters learned about drone usage through descriptions, images, or videos. However, the ability of the above presentations to narrate real-life scenarios is relatively limited. Firefighting operations can be more complicated, with numerous details under variable situations. How firefighters perceive the advantages and drawbacks of drone systems in actual operations remains unknown. Therefore, we conduct field trials to engage with firefighters in firefighting exercises. The goal is to evaluate our multi-drone system and collect insights about firefighters' perceptions under real cases.

\section{System description}

To support firefighters during an emergency response with autonomous drones, examining existing emergency response processes and firefighting challenges is essential. As part of the preliminary investigation, we first conducted and analyzed several interviews with representatives of various fire departments in Switzerland and studied the existing regulations. We collected data about the firefighting practices and gained insight into how they envision drones providing support. Additionally, the firefighters shared a firefighter handbook describing their procedures and instructions for handling diverse situations. Based on the gathered information, we can outline the initial firefighting scenario, encompassing the organizational structure of a firefighting team and the general response process to fire incidents. Next, we describe current practices, communication within a firefighter team, and the proposed drone system.

\subsection{Case description}

\begin{figure*}[]
\centering
\includegraphics[width=\textwidth]{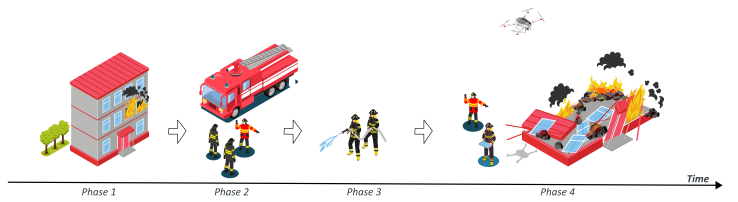}
\caption{Current firefighting procedure}
\label{fig:Timeline0}
\Description{A timeline showing four phases of a typical building fire operation. Phase 1: A building is on fire is reported. Phase 2: Firefighters arrive at the building. Phase 3: Firefighters extinguish fire with water pump. Phase 4: Firefighters check if there are still unextinguished fire covered under the building, sometimes with help from drones.}
\end{figure*}

Volunteer firefighting typically involves four major phases in response to an emergency (refer to Figure \ref{fig:Timeline0}):

\textit{Phase 1}: The firefighting operation commences with the receipt of a fire or emergency report via a phone call. Phase 1 encompasses the period from when the fire brigade receives the information to its arrival at the scene. During this phase, the commander is assumed to be the primary decision-maker. They must assess the fire situation, determine the required personnel and equipment for dispatch, and, if possible, establish an initial firefighting plan. Critical information such as terrain conditions, fire size, and available resources like water sources plays a pivotal role in decision-making during this phase.

\textit{Phase 2}: The second phase is the preparation stage upon reaching the scene. Here, the commander can provide a more precise assessment of the fire and task priorities while firefighters commence preparatory work.

\textit{Phase 3}: Subsequently, the firefighting phase entails the rescue and extinguishing of the fire. In the case of building fires, firefighters must make critical decisions, including whether and how to enter the building, among others. In wildfires, they need to determine the fire's extent, assess the risk of spreading, and identify the search area. If there are victims, firefighters evaluate their injuries and the complexity of the rescue operation to prioritize rescue efforts. One of the main challenges in this stage, especially in the context of large fires, is ascertaining the number and locations of injured individuals.

\textit{Phase 4}: The fourth phase pertains to firefighting activities after the fire has been extinguished. In the post-fire phase, the remaining tasks involve inspecting potential embers, documenting the site, and implementing any necessary remediation measures.

In many firefighting operations, one or two operational commanders take charge of overseeing the overall situation and making decisions regarding personnel and equipment deployment. When it comes to incorporating drone support selectively, the drone team operates under the authority of the commander. Moreover, stakeholders with the potential for information exchange during fire incidents include victims, bystanders, and medical personnel. Communication between firefighters and bystanders involves conveying information about the fire situation and evacuation instructions. Bystanders may also seek fire-related information. Once the firefighters have found the victims, firefighters must assess their physical condition and, on occasion, relay information to medical personnel. The information mentioned above, once filtered, is reported to the operational commander to formulate an overall situational assessment. Currently, the use of drones is initiated at the commander's request, and the drone team does not constitute an integral part of the on-scene fire unit.

\begin{figure}[]
\centering
\includegraphics[width=0.4\textwidth]{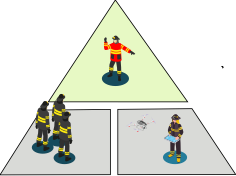}
\caption{Organization of the firefighting unit, including an optional drone team}
\label{fig:Hierarchy0}
\Description{A triangle containing three part to show the information flow within firefighting team. On top is the commander, receiving information from both firefighters in the lower left corner and drone officer in the lower right corner. Drone officer receives information from the tablet, which shows the interface of our drone system.}
\end{figure}

\begin{figure*}[]
\centering
\includegraphics[width=\textwidth]{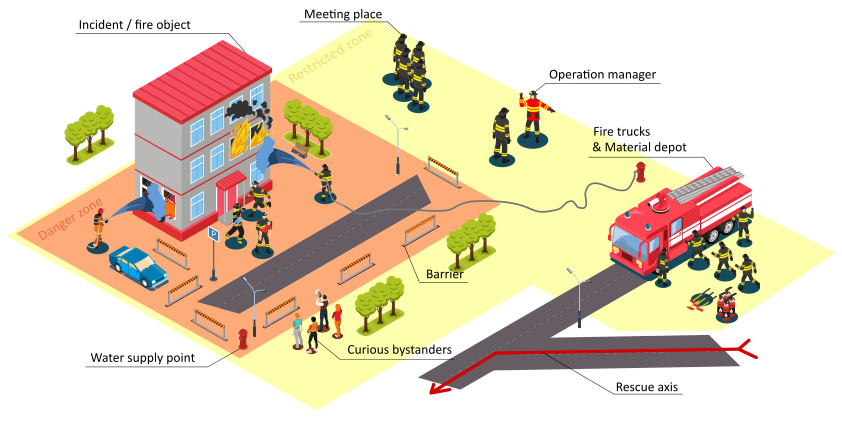}
\caption{Current firefighting scenario without autonomous drones}
\label{fig:scenarioWithoutDrones}
\Description{A figure simulating the building fire scene. The core area is called the danger zone, which includes the building on fire, firefighters who are trying to extinguish the fire, and the water supply. The danger zone is surrounded by barriers. Outside the danger zone is the restricted zone, which includes curious bystanders, the nearby road, fire trucks, other groups of firefighters, and the commander for this operation. The meeting zone for all firefighters is also in the restricted zone. This figure is to give an overview of what the typical building fire operation procedure and positions for different roles look like.}
\end{figure*}

The initiation of drone usage can occur at the commander's request during any phase. It's worth noting that the drone team is not an integrated component of the firefighting unit at the scene. Figure \ref{fig:scenarioWithoutDrones} illustrates a possible setup in a scenario without autonomous drones.

\subsection{Drone System} \label{Drone_system}

Over the past three years, we have worked on the use of drones in emergency services like police operations, gaining valuable insight and expertise in this dynamic field. We expanded the collected experience and design knowledge from existing literature through discussions with firefighters. Interviews with predefined questions helped us to generate ideas and suggestions for deploying the autonomous drone system and to expand the rationale with an empirical basis for design decisions. Based on the current firefighting operational procedures, we developed a scenario aimed at integrating autonomous drones into firefighting practices, including specific interactions between drones and firefighters or bystanders.

\begin{figure*}[]
\centering
\includegraphics[width=\textwidth]{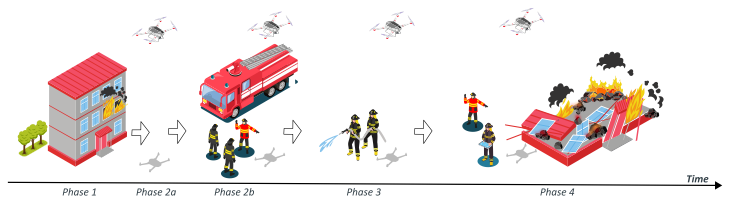}
\caption{Field trial firefighting procedure}
\label{fig:TimelineA}
\Description{A new timeline showing the phases of a building fire operation with the assistance of drones. Phase 1 and 4 remain the same as shown in Figure 2. Phase 2 is divided into two sub-phases. The whole figure is organized as follows. Phase 1: A building is on fire is reported. Phase 2a: Drones arrive at the building before firefighters arrive. Phase 2b: Firefighters arrive at the building, and check the situation based on information provided by drones. Phase 3: Firefighters extinguish the fire with a water pump, with a drone monitoring the scene in the air. Phase 4: Firefighters check if there is still unextinguished fire covered under the building, sometimes with help from drones.}
\end{figure*}

\begin{figure*}
\centering
\includegraphics[width=\textwidth]{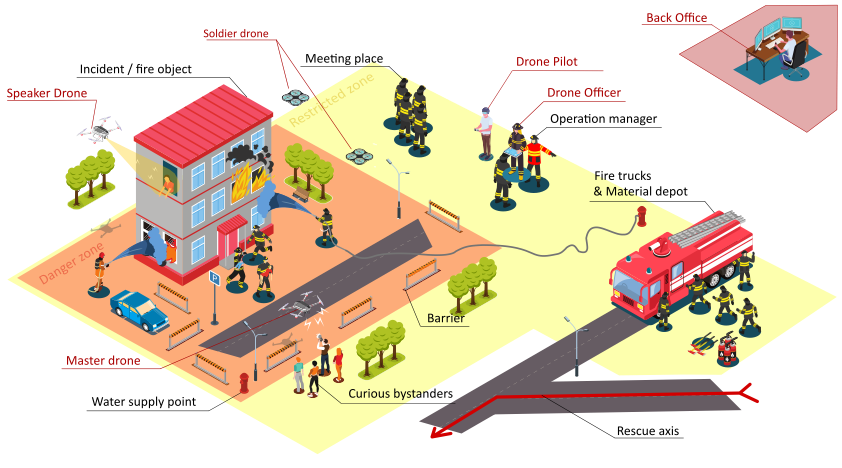}
\caption{Field trial firefighting scenario including autonomous drones}
\label{fig:scenarioDronesA}
\Description{A figure simulation the building fire scene with drones in the firefighting team. This figure is built based on Figure 3. The existing roles and their positions remain unchanged. Besides, we have the master drone, scouting above the fire scene for any accidents. And we have the soldier drones to deliver messages to the victims in the building and figure out what are their situation. A new character, drone officer, is added next to the commander to deal with information collected by drones and pass them to the commander. Also, the situation of the fire scene can be checked through the screen in the back office of the fire station.}
\end{figure*}

To fulfill the fire brigade commander's need for an initial on-site assessment, we introduced an autonomous drone right at the beginning of the operation. In our envisioned scenario, a preliminary scouting (master) drone departs from the fire station and arrives at the fire location ahead of the firefighting team (Phase 2a). An autonomous drone identifies potential fire points and victims and then dispatches the more agile soldier drone to confirm these findings. Figure \ref{fig:TimelineA} illustrates the suggested adaptations to the emergency operation process, while Figure \ref{fig:scenarioDronesA} depicts the modified organization of the fire scene.

\begin{figure*}[ht]
\centering
\includegraphics[width=\textwidth]{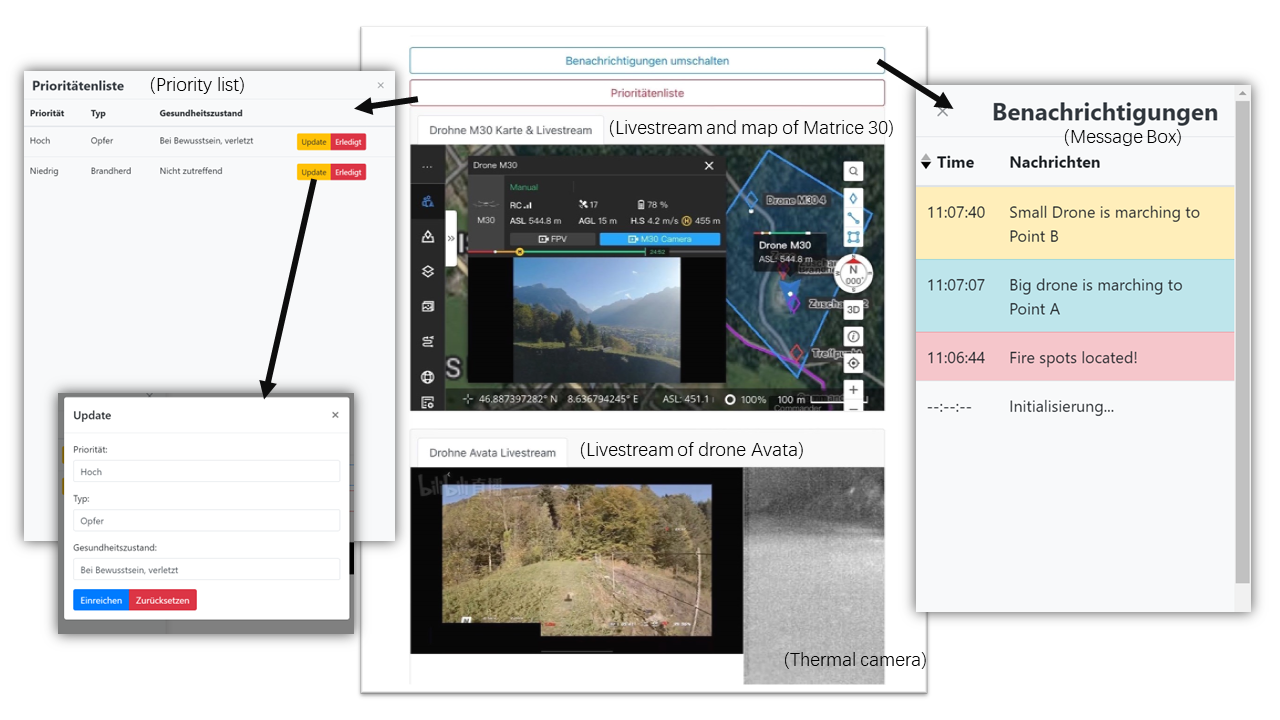}
\caption{Overview of the drone system user interface for Drone Officer in firefighting team}
\label{fig:Interface}
\Description{A figure showing the overview of the user interface and all the foldable components. On the interface, there are livestream and thermal camera videos of Avata, livestream, and map of Matrice 30, buttons for the message box and priority list. The message box component includes all the messages, sending time, and using different colors to mark the type and emergent level of the messages. The priority list component includes the type of incident, priority, and button to change the priority by hand.}
\end{figure*}

The proposed system employs two types of drones: the DJI Matrice 30, serving as the command (master) drone with night vision capability, the ability to operate in adverse weather conditions, and location reporting via Flighthub 2; and the DJI Avata, functioning as the soldier drone with motion control and a smaller form factor, allowing it to navigate indoor spaces. Both drones seamlessly interact with the system and can provide real-time services as required. Communication with the drone operator is facilitated through a tablet application supporting live streaming, an emergency case map, and decision-relevant notifications (refer to Figure \ref{fig:Interface}).

In designing human-drone interaction, our goal was to identify the most suitable interaction modes to facilitate effective communication between firefighters and drones. This selection was made thoughtfully, taking into account several critical factors, including the firefighters' familiarity with automation technology, information processing capabilities, and ability to respond swiftly while engaged in high-pressure tasks.

Informed by input from experienced firefighters, we prioritized audio and lighting interaction modes for effective communication during firefighting operations. The rationale behind opting for audio and lighting interaction modes is multifaceted. Firstly, these modes enable communication over extended distances, a vital aspect when dealing with potentially hazardous and expansive fire scenes. Moreover, they offer the advantage of minimal training requirements, making them accessible not only to firefighters but also to uninvolved individuals who may find themselves in emergencies. In our envisioned scenario, the drone first indicates who it is communicating with by rotating itself in the direction of the receiver and flashing its spotlight twice. The rotation of the drone on its own is not sufficient, as it is not always visible at great distances and in the dark. For audio/verbal communication, we decided to use audio broadcasting. Verbal addressing of specific people (e.g., with an exact description of the person) seems to be effective only to a certain extent. People who are not in the spotlight may ignore the message and not feel addressed as they prioritize the spotlight interaction mode. Conversely, people who are not addressed and are not in the spotlight may still feel involved in such situations, as people in emergencies tend to perceive all information as potentially important. Nonetheless, these different modes of interaction collectively serve to attract the attention of those in the immediate proximity of the emergency location.

However, it is essential to note that specific interaction modes were deemed unsuitable for our purposes. Gesture-based communication, while technologically appealing, presented drawbacks that outweighed its benefits. High latency and imprecise control \cite{tezza_state---art_2019} posed significant challenges, potentially exacerbating issues in larger fire scenarios. Furthermore, firefighters are often constrained in their ability to use gestures freely, rendering this mode impractical in many situations. Similarly, touch-based interaction and BCI were ruled out as viable options for communication between firefighters and drones. Safety considerations were pivotal in this decision, as both touch and brain-computer interaction necessitate physical contact or invasive procedures that could compromise safety in hazardous environments. Additionally, these modes exhibited slower input and output responses, a critical factor in emergency operations where rapid communication can be a matter of life or death.

\section{Method}
This study was part of a larger project that used a user-centered approach to determine the potential applications of autonomous drones to assist firefighters. We started with an ideation phase, after which we proposed an autonomous drone system. Subsequently, through two field trials conducted in collaboration with two fire departments, we aimed to evaluate firefighters' work processes, investigate the use of the drones to detect the source of the fire and possible victims, study the interaction with drones, and determine bystanders' perceptions. The primary objective of the field trials was to glean insights into how firefighters could effectively integrate autonomous drones into emergency operations, ascertain the optimal design principles for drone systems supporting firefighters, and understand the advantages and challenges perceived by citizens regarding drones that capture video footage of emergency situations and interact with victims and bystanders.

The field trials were part of regular firefighting exercises by two volunteer fire departments in two towns in Switzerland. In Switzerland, the spirit of volunteering is deeply ingrained, and in most cantons men and women dedicate themselves to the community on a part-time or voluntary basis. This commitment is not only reflected in the role of the militia in the army and national defense but also extends to civilian civil protection, especially the fire department. The seamless interaction between the military and fire departments is underlined by common ranks and rapid access to essential equipment and personnel, demonstrating a joint effort to protect the population and respond effectively to emergencies. The fire departments generally deal with all types of incidents. Their main tasks include firefighting, responding to natural disasters (e.g. hail, storms, snowfall, flooding), responding to reports from fire alarm systems, rescuing people on the road, engaging in operations on railroad systems, etc. Often two or more main tasks may overlap in one incident. During the fall 2022 exercises, we simulated a real firefighting operation in two distinct setups: a building fire and a forest fire, which we conducted with different participants. Firefighters were actively involved in realistic tasks and used prototype autonomous drones to manage emergencies. Subsequently, we conducted and recorded group discussions and individual interviews with the test subjects and used these recordings as the basis for our findings.

\subsection{Field trial scenario}

The field trial focused on fire incidents occurring within buildings or in forested areas and involved a total of 95 participants, categorized into three primary groups. First, we enlisted firefighters from two volunteer fire departments, where participants held positions as general commanders, exercise commanders, or civilian members of the firefighter department. Notably, these individuals were not career firefighters or part of the professional firefighting staff but rather dedicated volunteers contributing to firefighting efforts. Before the commencement of the field test, we established communication with the responsible exercise commander to delineate the conditions and exercise setup. While the primary commander's role was to oversee the entire operation, the exercise commander's task was to meticulously plan and organize the exercise to emulate the most realistic scenario. Consequently, exercise commanders were not directly involved in the firefighting or rescue operations.

In addition to the existing task forces, new roles were introduced. The general commanders received support from drone officers stationed alongside them. These drone officers accessed the drone’s video stream via a handheld digital device, evaluated the information, and subsequently relayed it to the commander. Participation in the exercises was compulsory for all civilian members of the department; albeit, they received limited information about the training schedule and the inclusion of drones in the operation.

Second, in accordance with the scenario, the field trial encompassed individuals with varying degrees of injury and mobility. The primary responsibility of these individuals was to remain stationed in assigned locations and await rescue. They were afforded the opportunity to interact realistically with their surroundings, including the drones, and collaborate with firefighters to aid in their location and rescue. This segment of the trial facilitated an exploration of the potential and challenges associated with drone interaction among different user groups.

\begin{figure}[]
\includegraphics[width=0.5\textwidth]{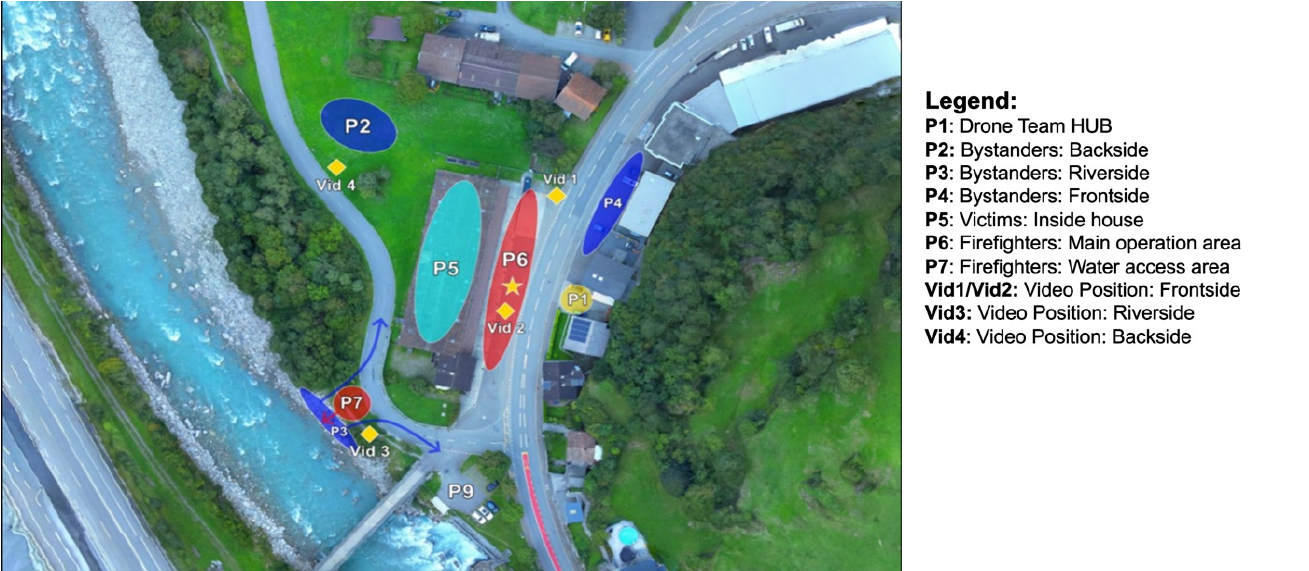}
\caption{The location used for the field trial with indicated relevant areas}
\label{fig:Location}
\Description{A bird-eye view of a building and its surroundings (grass fields, streets, river, trees) with indicated seven different areas: P1: drone team hub, P2: bystanders: backside, P3: bystanders: riverside, P4: bystanders: frontside, P5: victims, P6: firefighters: main operation area, P7: firefighters: water access area. Further, there are indicated four positions where cameras were placed to record the action of the firefighters and bystanders}
\end{figure}

Last but not least, the trial included subjects who were presumed to be indirectly affected by the fire, primarily serving as bystanders who happened to be near the fire scene. These participants were informed that they were taking part in a live fire exercise that would involve novel technologies. Bystanders were positioned at various locations, as depicted in Figure \ref{fig:Location}. To heighten the realism of the situation and simulate the varying attention levels of bystanders, all participants were assigned individual tasks unrelated to the fire incident. For instance, a small group was tasked with playing volleyball on the lawn at the backside of the building, thereby initially remaining unaware of the ongoing events or the use of drones. Throughout the field trial, bystanders were afforded the freedom to act according to their discretion. They could respond to requests from firefighters or drones as the situation unfolded. For documentation and evaluation purposes, the progress of the field test was recorded from different vantage points.

As the bystanders carried out their assigned tasks, the exercise commander initiated the field trial by either igniting the fire (in the case of a forest fire) or activating the smoke machine (in the case of a building fire). Within a few minutes, information regarding the fire incident reached the fire station, prompting the general commander to activate the alarm, resulting in the immediate mobilization of the entire crew. The command center promptly initiated the operation by issuing orders for the deployment of the drones. These drones departed from their base, located approximately 1500 meters away, and swiftly transmitted initial imagery of the emergency location. Merely one minute after the alarm was sounded, the drones were airborne. Meanwhile, it took the civilian members of the fire department an additional 15-20 minutes to arrive at the scene.

This operation involved the utilization of two drones operating as a coordinated team. While one drone provided an aerial overview and established communication with bystanders and potential victims through loudspeakers and spotlights, the second drone executed a thorough search for victims within the building or the forest. The operation concluded once all victims were successfully rescued, the fire was extinguished, and any potential residual embers were eliminated.

We seamlessly integrated the following preselected scenarios for human-drone interaction into the firefighters' routine exercises to replicate real-life conditions as accurately as possible:

\begin{enumerate}
 \item Verify the correct address and confirm the fire location
 \item Obtain information about the surroundings, including determining the extent of the fire
 \item Contact third parties and instruct them to vacate access roads, entrances, rescue lanes, and water resources
 \item Continuously gather information and evaluate future events
 \item Provide support and conduct a search for individuals, including potential victims, within the emergency area
 \item Conclude the operation and document incidents 
\end{enumerate}

To create the illusion of complete drone autonomy during our study, we employed a Wizard of Oz (WoZ) technique, following established methods \cite{karjalainen_social_2017, dahlback_wizard_1993, cauchard_drone_2015}. This approach enabled us to monitor and control the drone's actions to ensure the safety of all participants. By concealing the human operator behind a curtain, we crafted the impression among participants that the drone was functioning independently. This deliberate concealment facilitated an unbiased assessment of participants' interactions with the technology.

The field trials provided an effective means to evaluate the practical feasibility and effectiveness of autonomous drone systems in actual emergency situations. The selection of these scenarios was informed by an analysis of previous fire brigade operations, which revealed that the drone could reach the fire’s source more swiftly than the initial fire unit. Consequently, in the initial phase, the drone was often the sole representative of the firefighting team at the scene, raising questions about the drone’s legitimacy and the public’s perception of it.

\subsection{Participants and Data Collection}
A total of 67 firefighters (including commanders) participated in both exercises (see Table \ref{tab:participants}). The participation of the firefighters in the exercises was mandatory and was not remunerated. Some of the firefighters had prior experience using drones in emergency situations; they operated and used drones to investigate emergencies in the field. 

The research team employed the snowball method (word of mouth) and direct engagement with participating fire departments to recruit 22 volunteers for the bystander role. In recognition of their participation, which encompassed approximately 2 hours, bystanders received compensation amounting to 70 Swiss francs (comprising 25 Swiss francs per hour along with travel expenses) in adherence to university practices. These two hours encompassed preparation, active participation in the field trial, subsequent group and/or individual interviews, and scenario resolution. Furthermore, six victims participated in both field trials, and they were directly recruited from the participating fire departments.

Following the conclusion of each field trial, participants engaged in brief interviews conducted by a member of the research team. An internal fire brigade debriefing also took place after each exercise, with a representative from the research group permitted to attend these discussions. Due to the volunteer basis of the fire department, the exercises were scheduled in the evening to maximize firefighter participation. However, this timing posed a limitation for planned interviews, as not all participants were available for post-exercise interviews. Nevertheless, a total of 50 interviews were successfully conducted across both field trials (refer to Table \ref{tab:participants}). After Field Trial 1, interviews with bystanders averaged 12.7 minutes, while those with firefighters averaged 18 minutes. Subsequently, during Field Trial 2, interview durations increased, averaging 17.45 minutes for bystanders and 19 minutes for firefighters.

\begin{table} []
\caption{Number of subjects who took part in the field trial and number of interviews conducted per group of participants (in brackets)}
\label{tab:participants}
\begin{tabular}{lcc}
Participants group & Field Trial 1 & Field Trial 2 \\ \cline{1-3}
Bystander  & 11 (11)  & 11 (11)  \\
Victim  & 3 (0)  & 3 (3)  \\
Commander  & 1 (1)  & 1 (0)  \\
Exercise Commander & 2 (2)  & 2 (2)  \\
Drone officer & 1 (1)  & 1 (1)  \\
Firefighter  & 30 (7) & 29 (11) 
\end{tabular}
\end{table}

The interviews followed a semi-structured format, encompassing five primary topics: overall impressions, perceptions of drone performance, interactions with the drones, autonomy of the drones, and accountability and responsibility concerning drone actions. We delimited our focus to these topics for two primary reasons. First, expanding the scope of the interviews would have placed an excessive burden on the research team, and conducting immediate post-field trial interviews on a wider array of subjects would have been unfeasible. Secondly, we aimed to minimize waiting times to sustain participant motivation and engagement. For this study, our primary emphasis centers on questions related to overall impressions, perceptions of drone performance, interactions with the drones, and drone autonomy.

\subsection{Data analysis}
The primary objective of the field trial was to acquire insights into how autonomous drone systems should be designed to aid firefighters, how citizens perceive the interactions with drone systems, and what benefits and challenges drone systems could have on the firefighting team. The recorded interviews underwent transcription following the intelligent wording standard. Subsequently, the transcripts underwent analysis within the exploratory-interpretive paradigm \cite{saldana_coding_2021, stebbins_exploratory_2001}, wherein the researchers pursued a shared interpretation of the data rather than striving for absolute objectivity.

The data analysis unfolded in two distinct phases. First, a single researcher employed a bottom-up approach to identify predominant themes. This initial step led to the identification of 2399 pertinent passages, which were categorized under 51 secondary codes within five overarching themes: \textit{Firefighting Regulation}, \textit{Interaction}, \textit{Potential Applications}, \textit{Perceived Support} and \textit{Problem}. The sub-codes encompassed topics related to the suitability of drones in saving time, the efficiency of interaction, their impact on bystander emotions, issues of responsibility, moral decision-making, technical safety, transparency, and other pertinent considerations. For instance, sub-codes like "Situation awareness" and "Psychological effect" within the \textit{Perceived Support} theme were employed to categorize paragraphs discussing participants' comprehension of on-site situations based on drone-provided information and the comfort derived from the interaction. In Table \ref{tab:codes}, we present a description for every theme and some example sub-codes and data. The accuracy of coding underwent verification through iterative checks conducted by a second researcher, with borderline cases resolved through discussions.

In the second phase of analysis, the authors collaborated with fellow researchers in an iterative sense-making and data-restructuring process. This collaborative effort involved asynchronous exchanges through a shared repository and participation in two interpretation workshops, which included members of the authors' research group, including those who supported the field trial as assistants. Through this method, the authors systematically evaluated the comprehensiveness and authenticity of their interpretation, ensuring that the results were not solely shaped by individual perspectives but rather cultivated through an intersubjective approach. The sense-making process played a pivotal role in refining the organizational framework of our study's findings, leading to the consolidation of the initial five overarching themes into two primary meta-topics: enhancing autonomous drone systems and examining the impact of autonomous drone applications. This restructuring made it necessary to subject the previously coded data to a new structure. Specifically, the categories \textit{Perceived Support} and \textit{Problems} were combined into a comprehensive overarching category \textit{Impact on firefighters} which includes aspects such as changed processes, changes in team structure, training considerations and reliance dynamics. To provide a more nuanced understanding, the results were further subdivided based on the different emotions expressed in the statements. The explanation of these identified categories and their associated meanings is presented in the results section, taking into account the findings from the group discussions.

\section{Results}

\subsection{Improving autonomous drone system}

\subsubsection{Comments on current functions}
Since our research is centered around firefighters, we primarily assessed the sentiments and recommendations of the firefighters to evaluate these functions. However, to achieve a more comprehensive understanding of how these interaction methods can support firefighting activities in real-world scenarios, we also considered the assessments of other stakeholders in our analysis.

Participants from both trials acknowledged the effectiveness of both lighting and broadcasting interactions. The utility of the lights received positive feedback, as they facilitated victim location in dark and expansive areas. No dissenting opinions were voiced regarding this feature. Victims, bystanders, and firefighters involved in the rescue task also noted that the drone’s lighting improved visibility during nighttime operations. Besides, bystanders also showed affection for the spotlights as the lights increased the perceived credibility of drones and the information that came from them:

\begin{quote}
  F34\footnote{In our coding system, we employ distinct labels for various groups of participants to facilitate citations. Each label comprises a designation code followed by an ID number (\#): firefighter (F\#), commander (C\#), exercise commander (EC\#), drone officer (DO\#), victim (V\#), and bystander with (B\#) }:
  \itshape
  ``The rescue squad was then able to locate the drones using its positioning lights, so they were actually able to find these people relatively quickly.''
\end{quote}
\begin{quote}
  B13:
 \itshape
 ``And what I also saw was that they also pointed directly at things with the light beam. If you had to look at this on a map, you don't know if the location is here or a few trees further back. With the drone, you could see it well.''
\end{quote}

\begin{quote}
  B15:
 \itshape
 ``I have to say now, the big drone with the spotlight already seems pretty professional.''
\end{quote}

Participants held varying opinions regarding the interaction between drones and individuals over the radio. Some participants believed that the radio transmissions they received provided sufficient information about the situation and guided their subsequent actions. Specifically, firefighters found audio instructions valuable for evacuating uninjured bystanders from hazardous areas. Firefighters engaged in the evacuation task (F40) credited these instructions with effectively motivating people to move as required:

\begin{quote}
 F40:
 \itshape
 ``The drones can talk to them. That's what I've seen today, that they can make announcements or something. That they can address people.''
\end{quote}

Conversely, a different group of participants, primarily bystanders, questioned the effectiveness of radio broadcasts due to concerns about the optimal drone distance for audio communication. Two participants reported difficulty distinguishing the intended recipients of the instructions during drone broadcasts because the drone's position for audio transmission was often farther away from them than the distance between groups of participants. These participants also mentioned that the small drone could be bothersome and distracting when it flew too close to them. Therefore, designers should explore options for positioning the drone closer to participants for audio playback without causing resistance to the instructions or develop methods to differentiate audio broadcasts for various participant groups:

\begin{quote}
 B15:
 \itshape
 ``With the instructions, I also still see a little bit of disadvantages. Because there were two groups pretty close to each other. And there was a short moment where it was not quite clear to which group the instruction applied. And that can be very dangerous, of course. When the drone says: 'Go along the fence!' And there are two groups standing along the fence. Then both groups meet in the middle, and that may not be the point.''
\end{quote}

Furthermore, conflicting viewpoints emerged regarding the use of broadcasting for evacuation instructions and providing information about the fire situation to different participant groups. This discrepancy centered on the amount of information bystanders should be given during firefighting operations. According to the firefighting commander during the building fire trial, he preferred not to relay detailed information to crowds to avoid inducing fear and panic:

\begin{quote}
C1:
\itshape
 ``You (crowds) can see that there's a fire over there. But whether there's a trash can burning or something like that (should not be told to them). And I also don't want to have to tell them that there are still children in there or something like that. No, so that's good enough for me.'' 
\end{quote}

On the other hand, bystander groups expressed a desire for more information from drone broadcasts. They felt that they could not fully rely on simple instructions from drones directing them where to go, as they did not perceive drones as representing the firefighting agency. Their preferred interaction mode involved the drone providing on-site information so they could plan their next steps. However, they also recognized that relying on the drone for information reduced their need to approach firefighters for updates, which could be valuable during critical missions:

\begin{quote}
 B10:
 \itshape
 ``If you look from the outside, as it is an exercise, maybe all the processes should have gone in a bit more detail.''
\end{quote}
\begin{quote}
 B13:
 \itshape
 `` She just took the decision away from me because she just said, 'We're supposed to wait there for the fire department.' And maybe not going closer to the fire to watch, but yes wait for the fire department. It said something about a forest fire. We were told to wait there, that the fire department was on its way and there were not much more instructions than that.''
\end{quote} 

Feedback on the message and livestream functions was provided by two firefighters who directly interacted with the drone system in both trials. Information overload was a recurring concern discussed in relation to the drone system interface. The drone commander from the building fire trial emphasized the need to rapidly identify crucial messages while managing high-pressure tasks and the ability to review messages later. To address this issue, modifications were made for the second field trial, including changes to the message box and a priority list that reordered messages based on time, highlighting essential ones. Firefighters from the second trial expressed approval of the pop-up message feature when a victim was detected. They also suggested improvements to the livestream function, such as the addition of rewinding and enlarging capabilities to enhance the interface's comprehensibility.

\subsubsection{Design needs for next steps}
To explore potential interface improvements aligned with participants' needs, they offered several suggestions beyond the tested interactive experiences with the current drone system.

In line with the findings of Khan et al. \cite{khan_exploratory_2019}, participants expressed a desire for the drone system to detect hazardous incendiaries and investigate dangerous environments, particularly to benefit firefighters. Furthermore, the commander in the first trial discussed the potential of redirecting information to lower-level team members as a solution to address information overload. He proposed the idea of utilizing smart helmets equipped with real-time thermal camera images, marked assemblies, and environmental data collected by drones before firefighters enter the scene. Such implementation, from the commander's perspective, would alleviate the pressure associated with decision-making and task allocation.

If these smart helmets were to be integrated into the drone system, a crucial consideration would be the classification of information based on tasks and urgency levels before transmission to the interface.

Firefighters also expressed the need to integrate their current tools and documents, such as firefighting maps and digital systems, into the drone system. Both the commander and one of the firefighters from the second trial introduced the concept of merging these materials with the drone system:

\begin{quote}
 F38:
 \itshape
 ``We have an incident command system, and if you could couple that together a little bit, that would be great, of course. So, that the image runs in the back, you know?''
\end{quote}

\begin{quote}
 EC3:
 \itshape
 ``There's already a lot of predefined points… There I got water (pointing at a map)… There is a whole folder 'forest fire' where everything is stored… if you can program that into a drone so that the drone also has that information… then the drone can give me a smart suggestion then I say, 'Yeah, that's fine.' The rest is already defined.''
\end{quote}

Bystanders emphasized the importance of enhancing the credibility of drones for emergency evacuations. They highlighted that trust in a drone and its instructions hinges on the understanding that these autonomous vehicles are affiliated with a firefighting team. Participants suggested that proactive advertising through local media, such as radio, to inform the public about drones joining fire stations could help establish a connection between firefighters and drones. Additionally, participants recommended using distinctive markings on the drone's body, such as fire alarm symbols or blue lights, to convey its purpose for emergency response and build trust.

Interestingly, the firefighting commander's preference for drones to organize firefighters, rather than bystanders or victims, was unexpected. He explained that it's uncertain how individuals outside the firefighting team would respond to instructions from autonomous vehicles. This unpredictability could divert commanders' limited attention to episodic events and increase complexity, exacerbating the issue of information overload if interactions are not well-tested and simplified:

\begin{quote}
 C1:
 \itshape
 Yeah, you could see that nowadays, if we had an emergency and there would be bystanders, let’s say 100 people, and we would be saying: 'Please go to the side', 80 would have left immediately and said: 'Yes, yes. It's fine. Do your excellent job. We're leaving', and 20 would have still tried and come forward again somewhere else to look. And would have come there or there to look again.''
\end{quote}

Firefighters offered more specific feedback on the autonomous decision-making capabilities of drones and how they would interact with drones they perceived as autonomous. Many participants believed that semi-autonomous drones would be the most suitable solution for their current needs. In such cases, drones would facilitate the decision-making process by presenting potential solutions and optimal choices for humans to consider, with firefighters ultimately making the final decisions:
\begin{quote}
 EC3:
 \itshape
 ``If you can program that into a drone so that the drone also has that information. Then again it makes sense for it to say, 'Hey, in this case, operation commander, the location is there optimal or there optimal.' And then I also say, 'Yes, it is. It really is.'''
\end{quote}

Participants expressed mixed sentiments regarding the high level of autonomy in drone systems. On one hand, the two-way communication tested in the trials was deemed insufficient by both firefighters and bystanders. They expected more advanced communication capabilities, such as the ability to issue voice commands to visible drones or ask questions. Higher autonomy was seen as a means to provide intelligent objects for quicker, more complex two-way communication. On the other hand, many participants harbored reservations about completely autonomous drones or drones making decisions independently. Concerns centered on safety and privacy issues associated with autonomous drones. Interestingly, one firefighter even raised the concern that overreliance on drones could become a problem if drones operated independently without human supervision, potentially leading to humans neglecting crucial details while handling related tasks and losing control of the overall situation:

\begin{quote}
 F43:
 \itshape
 ``Yeah, I think it definitely would. I feel like if you always knew: 'The drone will be there.' I don't know, but I would imagine there would be a certain reliance on the device. And there would then also perhaps be a certain implicitness. People would say: 'Yeah, the drone went looking and it didn't find anything. It's fine.' So, then you would have to maybe take yourself by the nose a little bit and say. 'Hey, maybe go look again yourself.' ''
\end{quote}

Among these comments was the intriguing suggestion for additional training to familiarize drones with firefighting work, enabling them to find their place within the team due to their unique capabilities. Similarly, firefighters would benefit from training to understand new tasks and approaches when working with drones as new team members.

\subsection{Impact on Firefighting}

\subsubsection{Changes in the process}
Participants frequently cited time-saving as a prominent advantage of using drones. According to trial results, drones could reach the fire site 5-10 minutes faster than the firefighters. During this period, drones could survey the scene and relay information about the terrain and fire situation to the commander. Drones proved especially valuable for searching and reassuring victims in complex terrain or open spaces. According to one exercise commander, a key benefit was the potential rearrangement of task orders and decision-making locations with the inclusion of drones. Commanders could assign tasks before arriving at the scene:

\begin{quote}
 DO1:
 \itshape
 ``I mean, that was actually really positive. You actually had the information quickly and could accordingly send people to rescue or support there.''
\end{quote}
\begin{quote}
 F37:
 \itshape
 ``The best thing was that we actually found at least one person relatively quickly… It just sped up that we got to the places. So, it actually saved us time finding people.''
\end{quote}

\subsubsection{Changes in team structure}
In general, participants primarily viewed drones' roles during the trials as supportive tools for firefighters. For example, five participants described drones as a source or intermediary of information between the fire and firefighters. Several participants likened drones to teammates for firefighters. A firefighter from the building trial even remarked that the role of a drone felt more human-like than that of a tool. Similarly, a firefighter from the forest fire trial perceived drones as akin to co-workers capable of autonomous decision-making.

Based on preliminary interviews, many firefighters, particularly high-level commanders, expressed the need to filter information from drones to avoid information overload. The commander from the building fire trial proposed the idea of having a new team member alongside him to handle the video feeds and messages sent by drones during exercises. Another commander from the building fire trial suggested that the person responsible for managing drone-derived information should be an officer with expertise in drones and the operational process:

\begin{quote}
 DO2:
 \itshape
 ``That we say, I don’t always have to have people physically on-site, but we do it with a drone that can continuously fly over the entire fire object. And with thermal imaging, it actually needs an operator who can interpret the image, of course. That is, of course, crucial. ''
\end{quote}

\subsubsection{Changes in training}
Introducing a new team member into the firefighting unit would necessitate training to facilitate seamless collaboration. This experiment highlighted the fact that the majority of firefighters still lack a comprehensive understanding of drones' capabilities and limitations, leading to a lack of confidence when interacting with them. However, commanders expressed optimism about addressing this issue through repeated training and the integration of drone experts into the firefighting team:

\begin{quote}
 C1:
 \itshape
 ``You would have to practice it like we do it with everything else. You would have to use the drone in exercises like today and build on this and notice: 'What can the drone do?', 'What does the drone need to be able to do?', or: 'What are the drone’s limits?' So, without having practiced with it 10 times, of course, I would never take the drone into a mission. Otherwise, I really wouldn't know how to interact with it. So, you really have to practice it. Just like we have to practice now with the machines, that we have water. Practicing until you say: 'Now the drones are ready and I'm ready.'''
\end{quote}
\begin{quote}
 DO2:
 \itshape
 ``Because it was the first time, it was still very unfamiliar and not easy to understand all the equipment and how to use the interface. But it wasn't bad. It was just unfamiliar. So, it was not a problem. From my point of view, nothing was bad.''
\end{quote}

\subsubsection{Reliance}
Concerns also surfaced regarding the potential loss of control over firefighting operations due to reliance on information provided by drones. For several firefighters, decisions made by automated machines were not entirely trustworthy and could not serve as the sole source of information for building situational awareness on-site. Therefore, combining information from drones with data that firefighters could access and verify was deemed essential.

\begin{quote}
 F35:
 \itshape
 ``The only possible disadvantage I would see is that people were a bit too reliant or if they got information that they misinterpreted.''
\end{quote}

\section{Discussion}
\subsection{Choice of interactions}
Expanding upon Alon's previous research \cite{alon_drones_2021}, we have delved into alternative approaches for fostering direct communication between drones and firefighters, reducing reliance solely on commanders as intermediaries. Our exploration has encompassed the utilization of light signals and audio instructions as means of interaction. Our primary objective has been to gain insights into how firefighters perceive these novel interaction methods in real-world firefighting scenarios and to identify areas where potential improvements can be made. The findings from our study reveal that firefighters responded positively to the use of light signals as a form of guidance for locating victims. In contrast, the broadcasted audio instructions from the drones were met with some concerns regarding clarity.

One notable observation is that firefighters generally favor light-based interactions due to their simplicity and effectiveness. They say that information conveyed through light signals is straightforward and these signals remain visible in the sky until firefighters acknowledge them by reaching the designated location. This feature alleviates the need for immediate processing of information upon receipt, which stands as a notable advantage over audio interactions.

Based on these findings, we envision two potential avenues for enhancing these interaction methods. Firstly, visual interactions, facilitated by lights mounted on drones, could offer a more efficient means of communication for tasks beyond locating victims. Drawing inspiration from previous research on human-drone interaction \cite{e_drone_2017, ginosar_at_2023}, these lights could convey information about the current task, signal the next action, or issue commands through variations in color, flashing patterns, and blinking frequency. Within a firefighting context, it is plausible that drones could employ different light colors to indicate the urgency of a task, thus directly delivering critical information to firefighters without necessitating an intermediary commander. This would, of course, imply adequate training for the firefighters who need precise, ad hoc understanding of drones‘ light-based messages.

Secondly, enhancements can be made to audio interactions to ensure clarity and specificity, particularly when targeting specific groups. One potential solution involves utilizing light signals to illuminate the intended recipient group while simultaneously playing voice instructions. This dual-mode approach could potentially address concerns related to audio instructions lacking clarity and ensure that the information reaches the right audience effectively. Another issue about audio interaction is the perceived credibility. There are concerns about whether drones can be perceived as representative of firefighting teams and how this will affect trust from the perspective of emergency responders \cite{khan_exploratory_2019}. Our results show that bystanders have the same doubts when receiving information through broadcasting. In addition, Khan et al. also state that drones with prominent appearances for emergency services may increase perceived credibility \cite{khan_exploratory_2019}. According to bystanders in our field trials, they also prefer visual marks such as blue lights or firefighting symbols to a statement of affiliation before broadcasting. The potential reason could be that visual marks are more common for representing emergency response and can be constantly observed during operation.

\subsection{Information overload}
The issue of information overload in emergency response operations has been a longstanding concern, and the introduction of digital systems and devices has only exacerbated it \cite{jiang_ubiquitous_2004,cowlard_sensor_2010}. In line with the research conducted by Khan and Neustaedter \cite{khan_exploratory_2019}, it is evident that drone systems pose a significant challenge in terms of information overload. The autonomous drone system, at its current stage, has not entirely resolved this issue. Autonomous drones possess the capability to swiftly gather much information, yet this may introduce additional challenges. For instance, much information delivered at once might force the commander to make many decisions immediately. In that time, they would have limited cognitive resources to attend to information provided by firefighters. Therefore, exacerbation of information overload emerges as a primary challenge in the design of autonomous drone systems.
 
Previous literature has offered limited comprehensive solutions to this problem. However, our study yielded two potential measures proposed by firefighters to mitigate information overload in the scenario: enabling the autonomous drone system to provide selectable decision options and facilitating the storage of information in easily retrievable formats (such as video rewinding and message pop-ups). One notable concern regarding the first solution pertains to the credibility of decisions made by drones, especially in situations involving moral choices during rescue operations. Firefighters remain skeptical about autonomous drones making critical decisions based on essential information. Many of them still hold the belief that humans should be the decision-makers in such scenarios, as machines are perceived to lack empathy. In contrast, higher-ranking firefighters tend to see the advantages of machines, which can rapidly consider multiple factors and potentially exhibit a lower error rate than humans. Consequently, if the drone system were to assume the role of decision-maker, several aspects, including firefighter acceptance, model training processes, and success rates compared to humans, warrant further exploration.

Regarding the enhancement of information traceability, prior research has demonstrated the effectiveness of video rewinding in mitigating information overload, particularly in educational contexts \cite{oza_synchronous_2021}. Insights gained from interviews with firefighters reveal that specific firefighting tasks often adhere to established procedures. Such procedures, as well as local regulations, are embedded in files, maps, and the incident system. Those artifacts are also used to share information between firefighters and make decisions during operations. As of now, some of the artifacts are being used in physical form (maps), and others are digitized (incident system). This generates discontinuities between media. Adding drones amplifies this challenge, as drones provide very dynamic, rich information with higher temporal resolution. Therefore, the discussion on information traceability raises considerations on how to integrate existing data and systems with drone systems. Previous research suggests that the information provided by drones could help construct the mapping system for search and rescue \cite{alharthi_activity_2021}. Other researchers believe it is important to organize the current data before introducing a data stream from new communication approaches and build adjustable systems for different roles to select the information they need \cite{jones_remote_2020}. A later study suggests that live-streaming video should be used as a supplement to existing information sources rather than a replacement \cite{jones_rescuecastr_2022}. Our results indicate that the combination of previous and new data is also associated with the autonomy of the drones. An interesting comparison is that when firefighters suggested integrating drone information into the event system, they considered only the tool attributes of drone information collection. However, when the commander proposed having drones learn existing rules, they considered the possibility of drones making independent decisions. This insight can inspire interface designers of autonomous drone systems to pre-set a list of significant events that may occur during the process and annotate them in the video progress bar and message panel for easy reference in drone system design. The layout of interface modules, aimed at distributing information more logically, also emerges as a research topic deserving of further exploration.

Furthermore, our findings have highlighted another issue: when information is not adequately filtered and prioritized, drones may transmit irrelevant information to bystanders, potentially burdening firefighters with unnecessary work. This issue aligns with previous research findings \cite{van_erp_when_2015}. Some studies suggest that in emergency response situations involving interactions between responders and bystanders, the operation should focus on minimizing the negative impact of any conflicts rather than attempting to resolve them entirely \cite{van_erp_empowering_2018}. During the field trial, drones amplified bystanders' demands for more detailed information about the fire. This engenders controversy concerning the positive impact of drones on response times and may also contribute to exacerbating information overload because through the interaction with firefighters further irrelevant information might enter the exchange. One conceivable solution involves categorizing information permissions hierarchically and granting firefighters control over the level of information disseminated to bystanders. Previous research suggests that transparency about drones' use and their tasks might improve bystanders' sense of safety \cite{dolata_moving_2023}. It suggests that transparency about drones can be established prior to their use by firefighters through broad information campaigns instructing the population on how to behave and interpret the presence of drones. This could reduce the risk of conflicts and escalating information demand. Still, the psychological impact of drones on bystanders and the potential for them to escalate conflicts in extreme situations remain uncertain aspects that warrant further investigation.

\subsection{Change of configuration}
The alteration of firefighting configurations is another challenge for both drone system designers and firefighting teams. We can address discussions regarding this challenge from two perspectives: the role of drones after integration into the firefighting team and the potential change of existing firefighting roles.

Results from the field trials indicate that, at the current stage, most firefighters perceive drones as tools rather than companions. Firefighters primarily view drones as valuable for scouting and swiftly providing essential information. However, especially among those in higher-ranking roles, there is a belief that as drone autonomy advances, drones‘ acceptance and involvement in firefighting operations will increase. This suggests that the demand for drones to perform tasks with low levels of automation has not diminished with technological advancements. Instead, there is an emerging demand for drones to participate in decision-making processes. It’s conceivable that future drone swarm operations may involve autonomous drones in both decision-making and task execution, potentially forming hierarchical structures similar to human teams or even integrated in human chain of order. Unlike the findings in \cite{bjurling_drone_2020}, we anticipate that in the future, drones will become integral parts of the human team rather than separate entities comprising teams of drones and firefighters.

The inclusion of drones has highlighted the necessity for a role within the firefighting team that combines firefighting management skills with expertise in drone technology. Currently, it seems that a new commander would fill this role. However, as drones become more integrated into firefighting training, this role may eventually merge with the existing operation commander position. Additionally, many firefighters have already suggested that drones could effectively replace human counterparts in search and reconnaissance tasks, aligning with previous research findings. With a growing number of drones within the system, it's conceivable that the need for firefighters in reconnaissance tasks may decrease in the future. Firefighters’ focus may shift toward collaborating with or monitoring the automated decision-making processes of the drone system. Moreover, autonomous drone systems could alter the timing and location of decision-making, potentially allowing many decisions currently made on-site to occur before arrival or by personnel in a remote location. While this could reduce the risk of on-site accidents from negligence, it’s essential for the system to have a well-defined allocation of decision-making responsibilities and a designated person to avoid disputes \cite{jiang_ubiquitous_2004}.

\subsection{Communication demands of stakeholders}
Although our focus in this paper is on firefighters, comparing the viewpoints of other stakeholders with those of firefighters can provide insights into the design of drone systems. In search and rescue, communication is important to increase trust and reduce mental pressure for participants \cite{mencarini_becoming_2023}. Our results show that drone systems can facilitate communication among stakeholder groups. However, different groups have varying communication needs.

We see that firefighters, bystanders, and commanders have different attitudes toward sound interaction. They prefer lighting for different reasons: For firefighters and commanders, light improves visibility at night, which helps them to examine the scene and locate the victims. For bystanders, the spotlight increases the credibility of drones. For victims, light means they are identified and implies that rescue is coming soon, which provides comfort when they are trapped or injured. We believe this divergence arises from the distinct requirements of their roles. Firefighters consider the quality of interaction primarily based on its substantial assistance in specific firefighting tasks (evacuating the public, rescue operations, etc.). Preferred interactive methods should enable the drone to undertake part of the communication work with bystanders. High-ranked firefighters showed a focus on the simplicity of interaction and its potential impact. Their needs were driven by the immense workload and the psychological pressure of organizing and making decisions within a short time frame. Bystanders are more concerned about the manner and quality of interaction; they hope the interaction is friendly, reliable, and detailed. Victims seek more emotional support from interaction with drones. They wish the interaction to be more continuous and intelligent. 

Balancing the communication needs of different stakeholders will be a challenge. On the one hand, communication signals that can be commonly identified should be concise \cite{desjardins_collaboration_2014}, which aligns with the points of the commanders but for different reasons. Commanders wish to have clear but limited information passed to the public to avoid panic and overinterpretation of the situation. On the other hand, knowledge gaps during the crisis may contribute to the spread of rumors \cite{starbird_misinformation_2020}, indicating that passing on enough information to the public is necessary under emergencies. This aligns with firefighters' and bystanders' views. In addition to the previously discussed approach of controlling information flow through grading, determining the extent of information sharing based on the scenario and the stakeholders present may also be a solution. Another way to enhance communication is through repetitive training. Practicing with safety tools can optimize the coordination and processes of a rescue team \cite{desjardins_collaboration_2014}. Integrating drone systems into firefighter training not only builds trust in the system but also helps in defining the role of drones within the team. This facilitates the establishment of stable coordination methods and task processes.

Overall, the findings highlight the pervasiveness of information overload caused by drone systems across the design and implementation process. Whether in interface design, interaction mode selection, or structural reorganization, both firefighters and designers are likely to encounter the challenge of simplifying and directing information. If firefighters can widely accept autonomous drone decision-making, the system may reduce the volume of unprocessed information transmitted to them, thereby alleviating this issue. However, the application of autonomous drones in firefighting is still in its nascent stages, and many firefighters remain hesitant to trust decisions made by machines. This recalls the lessons from the Mann Gulch fire tragedy \cite{weick_collapse_1993}. Consequently, the process of integrating autonomous drones into firefighting decision-making must be iterative, allowing both firefighters and drones to adapt to each other’s presence. Incorporating human-in-the-loop training into the design process of autonomous drone systems, ensuring the involvement of all human team members, can be pivotal for future development and research.

\section{Limitation and Future Work}
The study primarily focused on the ways in which drone systems support firefighters during building and forest fires, offering valuable insights into these specific contexts. However, internal validity could be strengthened by expanding the investigation to include various firefighting tasks such as tunnel rescues, water rescues, fire monitoring, and predictions. Future research should broaden its scope to explore how drones can effectively aid in these diverse tasks, providing a comprehensive understanding of their potential applications and impact. Another area for future investigation lies in exploring the quantitative impact of larger drone swarms on firefighting and rescue operations. Our study concentrated on a restricted set of interaction methods based on light, voice, and interface support, tailored to firefighters' needs during the trial. Subsequent research could examine the effectiveness and feasibility of incorporating voice to specifically address receivers, touch, and brain-computer interfaces in real firefighting scenarios, offering insights into novel ways of enhancing communication and control.

A limitation of the study relates to the sample size, as perspectives were gathered from only two voluntary fire departments in Switzerland. To enhance external validity, future research should include a more extensive sample of field trials that encompasses a broader range of fire departments, both volunteer and professional. The study captured the perspectives of only two voluntary fire departments in Switzerland, and the transferability of results to professional departments may be nuanced. Aspects such as personnel availability, particularly qualified unmanned aerial vehicle (UAV) commanders, might be perceived differently among professional firefighters compared to their volunteer counterparts. This approach would provide a more representative understanding of how drone systems could be applied in different firefighting contexts. Additionally, our research was geographically confined to Switzerland, and the influence of local firefighting regulations and cultural factors on participants' perspectives may have introduced biases. Expanding the study to encompass diverse firefighting contexts globally would enhance the generalizability and applicability of our findings.

\section{Conclusion}
This research explored the functions and impact of the autonomous drone system through two field trials, simulating building and forest fire operations, respectively. We conducted field trials to simulate the autonomous features of the system and gathered data from participants to assess their perceptions and offer suggestions for such a system. The results pertaining to light, sound, and interface interactions can serve as guidance for drone system designers. Light is expected to find more applications in future designs, while voice interactions have the potential to facilitate firefighters' tasks while also raising concerns related to privacy and unforeseen events. Additionally, our findings reveal the challenges associated with addressing information overload and adapting to changes in team structure as firefighters incorporate autonomous drones into their existing configurations. Suggestions for mitigating information overload can benefit both designers and firefighters, aiding in the development and utilization of these systems in firefighting contexts. Furthermore, our findings on the impact of autonomous drone systems suggest that these systems could entail increased responsibilities for making decisions related to drone operations, offering insights into potential role reorganization within firefighting teams. Our study contributes to the HCI community in terms of drone system design and also provides suggestions for the firefighting teams for future changes.

\begin{acks}
This study was conducted as part of the project 'Automated Decision-Making and the Foundations of Political Authority,' funded by the Swiss National Science Foundation (project number 208013). We express our sincere gratitude to all project members for their valuable feedback and active involvement during the development and evaluation of the solution. Special thanks to the participants in the field trials for generously sharing their opinions and feedback. Our appreciation also goes to the fire stations in Altdorf and Silenen, as well as the firefighters in the trials, for their unwavering support and collaboration during the exercises.

We extend our thanks to the members of the University of Zurich for their crucial support during data collection and analysis: Yiwei Wu, Andreas Bucher, Sven Eckhardt, Beat Furrer, Simon Giesch, Yiming Hu, Yinglun Liu, and Pascal Marty. Additionally, we express our gratitude to the anonymous participants of our study and the review team for their invaluable advice concerning this manuscript.
\end{acks}

\bibliographystyle{ACM-Reference-Format}
\bibliography{references}


\appendix

\section{Description of codes, sub-codes and example of data}

\label{tab:codes}
\begin{tabular}{|c|c|c|l|} 
\hline
\textbf{Codes} & \textbf{Description} & \textbf{\begin{tabular}[c]{@{}c@{}}Examples of \\ subcodes\end{tabular}} & \multicolumn{1}{c|}{\textbf{Example of data}}                                                           \\ \hline
\multirow{3}{*}{\begin{tabular}[c]{@{}c@{}}Firefighting \\ Regulation\end{tabular}} & \multirow{3}{*}{\begin{tabular}[c]{@{}c@{}}Firefighting incidents \\ and corresponding \\ measures.\end{tabular}}         & Type of fire                & \textit{\begin{tabular}[c]{@{}l@{}}… there is the room fire, which is the next biggest type \\ of fire, and is already very critical. And beyond that, \\ we have fires in commercial buildings, fires in entire \\ residential buildings, industrial fires, and so on … \end{tabular}} \\ \cline{3-4} 
 &  & Organization & \textit{\begin{tabular}[c]{@{}l@{}}… for example, how the damage site is organized. We \\ have a material depot, we have an assembly area, we \\ have the incident, and we have the operation \\ commander's location … \end{tabular}} \\ \cline{3-4} 
 &  & Communication & \textit{\begin{tabular}[c]{@{}l@{}}… we had radios and we were connected to each other \\ like that, and the exercise commander was standing \\ next to me … \end{tabular}} \\ \hline
\multirow{3}{*}{Interaction} & \multirow{3}{*}{\begin{tabular}[c]{@{}c@{}}Interactions participants \\ captured in the trials and \\ how they feel about them.\end{tabular}}  & Livestreaming  & \textit{\begin{tabular}[c]{@{}l@{}}… so that interaction, I saw where the drone was, \\ what it was doing, and where the injured persons\\ were … \end{tabular}} \\ \cline{3-4} 
 &  & Lightening  & \textit{\begin{tabular}[c]{@{}l@{}}… I also saw that they pointed directly at things with \\ light. If you have to look at the map, you don't know \\ where the location is as there are a few trees behind: \\ with drones, you can see it well … \end{tabular}} \\ \cline{3-4} 
 &  & Audio   & \textit{\begin{tabular}[c]{@{}l@{}}… the information came to me via audio, I was able \\ to brief the rescue team accordingly … \end{tabular}} \\ \hline
\multirow{3}{*}{Support}  & \multirow{3}{*}{\begin{tabular}[c]{@{}c@{}}Perceived support from \\ the drone system during \\ the field trials.\end{tabular}}   & Situation awareness  & \textit{\begin{tabular}[c]{@{}l@{}}… so I saw that someone needed to be rescued and \\ was able to pass it on to the command center. That \\ was actually the positive (aspect) of the whole thing … \end{tabular}} \\ \cline{3-4} 
 &  & Victim rescue & \textit{\begin{tabular}[c]{@{}l@{}}… so I was in the rescue troop. That's why we had\\ to deal with the drone. And that was actually \\ great that you had a clue where the person was … \end{tabular}} \\ \cline{3-4} 
 &  & Instructions  & \textit{\begin{tabular}[c]{@{}l@{}}… there was a drone broadcasting to us, saying that \\ us need to evacuate from here since the building \\ is on fire. It can be dangerous. I think its helpful … \end{tabular}}  \\ \hline
\multirow{3}{*}{Application} & \multirow{3}{*}{\begin{tabular}[c]{@{}c@{}}Suggestions regarding new\\ feature in the system or \\ improvement of existing\\ features.\end{tabular}} & \begin{tabular}[c]{@{}c@{}}Suggest optimal \\ decision\end{tabular}  & \textit{\begin{tabular}[c]{@{}l@{}}… It can make suggestions, which would be really\\ helpful. I really think that can take some burden \\ of the commander … \end{tabular}} \\ \cline{3-4} 
  &  & \begin{tabular}[c]{@{}c@{}}After-fire\\ check\end{tabular} & \textit{\begin{tabular}[c]{@{}l@{}}… an ember, which is then covered with black ash. \\ You see from the outside, you think: "Yes, it's \\ extinguished," but then you have to check with \\ the thermal imaging camera to be sure … \end{tabular}} \\ \cline{3-4} 
  &  & Delivery  & \textit{\begin{tabular}[c]{@{}l@{}}… I know there are drones that can lift a lot of things \\ as well. So, you might be able to provide something \\ to a person somewhere that's in a hardly accessible \\ location even if it's just a water bottle … \end{tabular}}  \\ \hline

\multirow{3}{*}{Problem}  & \multirow{3}{*}{\begin{tabular}[c]{@{}c@{}}Perceived problems or \\ potential concerns about\\ using drones for firefighting\end{tabular}} & Privacy  & \textit{\begin{tabular}[c]{@{}l@{}}… data protection is certainly an issue. Personal \\ security, so that it is simply there … \end{tabular}}   \\ \cline{3-4} 
  &   & Technical  & \textit{\begin{tabular}[c]{@{}l@{}} … a technical limit that was reached, but it's a \\ technical system and every technical system has \\ its technical limit … \end{tabular}}                            \\ \cline{3-4} 
  &   & Interaction & \textit{\begin{tabular}[c]{@{}l@{}}… the radio was going non-stop, of course. So, \\ there was just too much information … \end{tabular}} \\ \hline
\end{tabular}

\end{document}